\documentclass[preprint,review, 12pt]{article}
\usepackage{algorithmicx}
\usepackage{algorithm}
\usepackage{algpseudocode}
\usepackage[utf8]{inputenc}
\usepackage{amsmath,amsthm,amsfonts,amssymb,amscd}
\usepackage{lineno,hyperref}
\usepackage{float}
\usepackage{multirow,booktabs}
\usepackage{amsmath}
\usepackage{algorithm}
\usepackage{subcaption}
\usepackage{amssymb}
\usepackage{amsmath} 
\usepackage{bm} 
\usepackage{mathtools} 
\usepackage{mathrsfs}
\usepackage{graphicx}
\usepackage[x11names]{xcolor}
\usepackage[utf8]{inputenc}
\usepackage[T1]{fontenc}
\usepackage[margin=2cm]{geometry}

\usepackage[lighttt]{lmodern}

\makeatletter
\algrenewcommand\ALG@beginalgorithmic{\ttfamily}
\makeatother

\usepackage{caption}

\newlength\myindent
\title{KF-PLS: Optimizing Kernel Partial Least-Squares (K-PLS) with Kernel Flows}
\author{
  Zina-Sabrina Duma\textsuperscript{1}\textsuperscript{*} \and 
  Jouni Susiluoto\textsuperscript{2} \and 
  Otto Lamminpää\textsuperscript{2} \and 
  Tuomas Sihvonen\textsuperscript{1} \and 
  Satu-Pia Reinikainen\textsuperscript{1} \and 
  Heikki Haario\textsuperscript{1}
}

\begin{document}

\maketitle

\footnotetext[1]{LUT University, Yliopistonkatu 34, Lappeenranta 53850, Finland, *Zina-Sabrina.Duma@lut.fi}
\footnotetext[2]{Jet Propulsion Laboratory, California Institute of Technology, 4800 Oak Grove Drive, La Cañada Flintridge, CA 91011, USA}

\begin{abstract}
    Partial Least-Squares (PLS) Regression is a widely used tool in chemometrics for performing multivariate regression. PLS is a bi-linear method that has a limited capacity of modelling non-linear relations between the predictor variables and the response. Kernel PLS (K-PLS) has been introduced for modelling non-linear predictor-response relations. In K-PLS, the input data is mapped via a kernel function to a Reproducing Kernel Hilbert space (RKH), where the dependencies between the response and the input matrix are assumed to be linear. K-PLS is performed in the RKH space between the kernel matrix and the dependent variable. Most available studies use fixed kernel parameters. Only a few studies have been conducted on optimizing the kernel parameters for K-PLS. In this article, we propose a methodology for the kernel function optimization based on Kernel Flows (KF), a technique developed for Gaussian process regression (GPR). The results are illustrated with four case studies. The case studies represent both numerical examples and real data used in classification and regression tasks. K-PLS optimized with KF, called KF-PLS in this study, is shown to yield good results in all illustrated scenarios. The paper presents cross-validation studies and hyperparameter analysis of the KF methodology when applied to K-PLS.
\end{abstract}

\section{Introduction}

Partial Least-Squares (PLS) is an iterative method for finding latent variables (LVs) that maximize covariance between the input and response variables \cite{Helland2014}. LVs represent linear combinations of the input variables \cite{Burnham1996}. PLS is widely used in spectral analysis as it compresses the useful information from numerous bands into a low number of LVs. The LVs left out include, for example, background variation or variation not related to the observed response \cite{Kawamura2019}. PLS has been used in a wide range of applications such as neuroimaging \cite{Krishnan2011}, linguistics \cite{Hair2022}, hydrology \cite{Rajbanshi2020}, food sciences\cite{RasooliSharabiani2022} and remote sensing \cite{MeachamHensold2019}.

PLS is a bi-linear regression method, with limited capacity to model non-linear phenomena \cite{martens2001analysis}. In many applications, the dependencies between the predictor and response variables might be non-linear. Such applications include, for example, estimating the crack length of steel as a function of chemical composition, mechanical properties and welding parameters \cite{lin2023generating}, modelling river phosphorus dynamics \cite{timis2022advancing}, NIR calibration, where scattered light or absorption band non-linearities exist \cite{blanco2000nir}. 

In Kernel PLS (K-PLS), the predictor variables are mapped to a Reproducing Kernel Hilbert (RKH) space, which is usually higher-dimensional, where the relationship between the kernel matrix and the dependent variable is linear; thus, PLS can be justifiably applied\cite{Wang2015}. K-PLS been so far applied to, for example, modelling fault detection in chemical systems \cite{fazai2019online}, process monitoring based on key performance indicators \cite{si2020key}, monitoring wastewater treatment \cite{liu2021monitoring},  image super-resolution \cite{li2015kpls}, and many other domains. 

To estimate the parameters of a given kernel, existing methodologies include the kernel target alignment \cite{Yin2012}, feature space matrix \cite{Nguyen2008}, largest variance criteria \cite{Yang2009}, dimension and variance \cite{Kim2005}, sample distance \cite{Kenig2010}, maximum distance to mean \cite{Teixeira2008}, sum of variable spread \cite{Ni2011}, and genetic algorithm \cite{MelloRomn2020}. However, in most case studies, the kernel function is Gaussian \cite{Wang2015}, and the kernel parameter is set to a constant value or found through line search \cite{rosipal2001kernel}.

In Kernel Flows, the kernel function is learned from the data by minimizing a loss function that is obtained through cross-validation \cite{owhadi2019kernel}. The present article proposes using the Kernel Flows methodology to optimise K-PLS regression, select the kernel function, and learn the parameters. In addition, this work presents the option of using a combination of kernel functions.

The methodology was tested with four case studies. The case studies include simulated and real-world data. The studies represent non-linear regression and classification tasks. The KF-PLS results are compared with regular PLS, K-PLS without any optimization, and K-PLS optimized by using other methodologies than Kernel Flows. The other optimizing methodologies include Genetic Algorithm and Nelder-Mead simplex. The performance of the classification tasks has been evaluated in classification accuracy metrics, whereas the  success in the regression tasks  has been measured in root mean square error (RMSE) metrics. Other indicators of performance, such as convergence time and run stability, are considered. 

Our proposed method is novel in several ways. First, K-PLS is computed with the SIMPLS \cite{de1993simpls} approach, and the kernel matrix deflation varies from the original proposal \cite{rosipal2001kernel}. Second, the Kernel function parameters are learned using Kernel Flows methodology, which has not been previously applied to K-PLS. Third, a combination of scaled kernels has been introduced to K-PLS.  

\section{Mathematical Methods}
In this section, we present the mathematical methods used for optimizing K-PLS with KF. The PLS methodology is described in Sec. \ref{ssec:PLS}, whereas the kernel version of PLS is showcased in Sec. \ref{ssec:KPLS}. The KF methodology is presented in Sec. \ref{ssec:KPLS-KF}. The data pre-treatment and modelling approach is presented in Sec. \ref{ssec:pretreatment}.

\subsection{Partial Least-Squares Regression}\label{ssec:PLS}

The main PLS algorithms are: Non-linear Iterative Partial Least Squares (NIPALS) \cite{wold1975soft}, SIMPLS \cite{de1993simpls}, and Orthogonal PLS (O-PLS) \cite{verron2004some}. Originally, K-PLS used the NIPALS-PLS algorithm \cite{rosipal2001kernel}. The PLS algorithm used in this study, SIMPLS, is presented in Algorithm \ref{alg:PLS} below. SIMPLS is faster to compute and requires less memory than NIPALS \cite{de1993simpls}. Principal Component Analysis (PCA) \cite{Bro2014} is used inside the SIMPLS algorithm for covariance matrix decomposition. The matrices of interest outputted by the PLS algorithm used in K-PLS are $\mathbf{W}$, $\mathbf{P}$ and $\mathbf{Q}$. The matrix $\mathbf{W}$ represents the $x$-side loadings, rotated in the direction of maximum covariance between the input data matrix $\mathbf{X}$ and the output data matrix $\mathbf{Y}$. The $\mathbf{P}$ matrix contains the un-rotated $x$-side loadings, and the $\mathbf{Q}$ matrix contains the loading vectors for the $\mathbf{Y}$ response side. The number of vectors in these matrices equals the number of LVs selected in the model. 

\begin{algorithm}[H]
\caption{SIMPLS Algorithm - \textit{pls}($\mathbf{X}$, $\mathbf{Y}$, $n_{LVs}$)}\label{alg:PLS}
\textbf{\textit{Input}}: input matrix ($\mathbf{X}$), response ($\mathbf{Y}$), number of latent variables ($n_{LVs}$).\textbf{\textit{Output}}: X-Side score matrix ($\mathbf{T}$), x-side loading matrix ($\mathbf{P}$), x-side rotated loadings matrix (weights) ($\mathbf{W}$), y-side score matrix ($\mathbf{U}$), y-side loadings matrix ($\mathbf{Q}$).
\begin{algorithmic}[1]
\State $    \mathbf{C} \gets \mathbf{Y}^T \mathbf{X}$ \Comment{Calculate the covariance between $\mathbf{X}$ and $\mathbf{Y}$ matrices.}
\For{$i \gets 1$ to $n_{LVs}$}
\State $   \mathbf{w} \gets pca(\mathbf{C}) $ \Comment{Extract the PCA loadings from the first PC in a column vector $\mathbf{w}$.}  
\State $    \mathbf{t} \gets \mathbf{X} \mathbf{w}$     \Comment{Calculate $\mathbf{X}$-side scores for $LV_i$.}
\State $    \mathbf{q} \gets  \frac{\mathbf{Y}^T \mathbf{t}}{\mathbf{t}^T \mathbf{t}} $ \Comment{Calculate $\mathbf{Y}$-side loadings for $LV_i$.}
\State $    \mathbf{u} \gets \frac{\mathbf{Y} \mathbf{q}}{\mathbf{q}^T \mathbf{q}} $ \Comment{Calculate $\mathbf{Y}$-side scores for $LV_i$.}
\State $    \mathbf{p} \gets  \frac{\mathbf{X}^T \mathbf{t}}{\mathbf{t}^T \mathbf{t}} $   \Comment{Calculate $\mathbf{X}$-side loadings for $LV_i$.}
\State Store $\mathbf{t}, \mathbf{p}, \mathbf{q}, \mathbf{u}, \mathbf{w}$ into $\mathbf{T}, \mathbf{P}, \mathbf{Q}, \mathbf{U}, \mathbf{W}$
\State $   \mathbf{C} \gets \mathbf{C}-\mathbf{P}(\mathbf{P}^T\mathbf{P})^{-1}\mathbf{P}^T\mathbf{C}$ \Comment{Deflate $\mathbf{C}$ matrix.}

\EndFor
\end{algorithmic}
\end{algorithm}
In classification, the $\mathbf{Y}$ matrix will contain as many columns as there are classes. Every response variable will respond to the question \textit{Is the sample member of the class?}, with 1 confirming membership and 0 denying membership. Each observation will have the value '1' in the response variable of its class and 0 to all the others. The procedure is known as Partial Least-Squares with Discriminant Analysis (PLS-DA). The kernelized version of the method will be denoted in this paper with K-PLS-DA.

\subsection{Kernel functions}\label{ssec:functions}

The mapping of the input data into RKH space is done via the kernel trick \cite{scholkopf2000kernel}. This paper explores the Gaussian, Cauchy and Matern kernels and a combination of kernels.

The Gaussian kernel is given by
\begin{equation}
    k_G(\mathbf{x}, \mathbf{y}) =  exp \left( - \frac{|| \mathbf{x} - \mathbf{y} || ^ 2}{2 \sigma^2} \right).
    \label{eq:gaussianKernel}
\end{equation}

The Matern $ \frac{1}{2} $ or Laplacian kernel, Matern $ \frac{3}{2} $ kernel, and the Matern $ \frac{5}{2} $ kernel are given by the respective formulas below

\begin{align} 
    k_{M1/2}(\mathbf{x}, \mathbf{y}) &= exp \left(- \frac{|| \mathbf{x} - \mathbf{y} ||}{\sigma} \right) \label{eq:matern1/2}\\ 
    k_{M3/2}(\mathbf{x}, \mathbf{y}) &= \left(1 + \frac{\sqrt{3}|| \mathbf{x} - \mathbf{y} ||}{\sigma}\right) exp\left(- \frac{\sqrt{3}|| \mathbf{x} - \mathbf{y} ||}{\sigma}\right) \label{eq:matern3/2}\\
    k_{M5/2}(\mathbf{x}, \mathbf{y}) &= \left(1 + \frac{\sqrt{5}|| \mathbf{x} - \mathbf{y} ||}{\sigma} + \frac{5|| \mathbf{x} - \mathbf{y} ||^2}{3\sigma^2}\right) exp\left(- \frac{\sqrt{5}|| \mathbf{x} - \mathbf{y} ||}{\sigma}\right). \label{eq:matern5/2}
\end{align}

The Cauchy kernel is written as:

\begin{equation}
    k_C(\mathbf{x}, \mathbf{y}) = \frac{1}{1 + \frac{|| \mathbf{x} - \mathbf{y} ||^2}{\sigma^2}}.
    \label{eq:cauchy}
\end{equation} 

A regularization parameter $ \delta$ can be added to the Kernel matrix

\begin{equation}
    \mathbf{K}_r = \mathbf{K} + \delta \mathbf{I}.
    \label{eq:regularizaion}
\end{equation}

If additive kernels are used, a parameter that regulates the contribution of each kernel is added for each of the mapping functions,

\begin{equation}
    \mathbf{K}_f = \gamma_1 \mathbf{K}_G + \gamma_2 \mathbf{K}_{M1/2} + \gamma_3 \mathbf{K}_{M3/2} + \gamma_4 \mathbf{K}_{M5/2} + \gamma_5 \mathbf{K}_C +  \delta \mathbf{I}.
    \label{eq:family}
\end{equation}

\subsection{Kernel Partial Least-Squares}\label{ssec:KPLS}

Rosipal and Trejo introduced K-PLS, 2001 \cite{rosipal2001kernel}, and the detailed methodology is available in multiple sources \cite{rosipal2001kernel}, \cite{Wang2015}. The algorithm used in the original source is K-PLS-NIPALS, whereas in this paper K-PLS-SIMPLS algorithm is used. Algorithm~\ref{alg:KPLS} presents how the kernel matrix is calculated to input into the PLS algorithm to yield regression coefficients ($\mathbf{B}$). These regression coefficients can then be utilized to make predictions on new data or validate the model with newly-acquired or test data ($\mathbf{X}_{test}$).

\begin{algorithm}[H]
\caption{Obtaining regression coefficients by K-PLS - $kpls(\mathbf{X}, \mathbf{Y}, n_{LVs}, \boldsymbol{\theta})$}\label{alg:KPLS}
\textbf{\textit{Data}}: training predictor matrix ($\mathbf{X}$), response variable ($\mathbf{Y}$), number of PLS dimensions ($n_{LVs}$), kernel function parameters  ($\boldsymbol{\theta}$). \textbf{\textit{Output}}: regression coefficients ($\mathbf{B}$) 
\begin{algorithmic}[1]
\State $\mathbf{K} \gets k(\mathbf{X},\mathbf{X}, \theta) $  \Comment{Map training data $\mathbf{X}$ using kernel function.}
\State $\tilde{\mathbf{K}} \gets (\mathbf{I} - \frac{1}{n}\mathbf{1}_n\mathbf{1}_n^T)\mathbf{K}(\mathbf{I} - \frac{1}{n}\mathbf{1}_n\mathbf{1}_n^T)$ \Comment{Center kernel matrix.}
\State $\mathbf{W, P, Q} \gets pls(\tilde{\mathbf{K}}, \mathbf{Y}, n_{LVs})$ \Comment{Compute \textit{PLS} for a number of LVs [See Algorithm \ref{alg:PLS}].}
\State 
$ \mathbf{B}  \gets \mathbf{W} (\mathbf{P}^T \mathbf{W})^{-1} \mathbf{Q}^T $  \Comment{Calculate regression coefficients.}
\end{algorithmic}
\end{algorithm}

In Algorithm \ref{alg:KPLS}, $\mathbf{1}_n$ is a vector with the size of the number of observations data points in $\mathbf{X}$. Each of the vector's values is one divided by the number of data points ($1/n$). In the same manner, $\mathbf{1}_{nt}$  below is a vector with the number of elements equal to the number of data points in $\mathbf{X}_{test}$. $\mathbf{I}$ represents the identity matrix.

To make predictions for newly acquired data (denoted with the \textit{test} subscript) with the regression coefficients, one has to map the new data ($\mathbf{X}_{test}$) using the kernel function $\mathbf{K}_{test} = k(\mathbf{X}_{test},\mathbf{X}, \boldsymbol{\theta})$. The new kernel matrix has to be centred according to the training kernel centres \cite{rosipal2001kernel}

\begin{equation}
    \tilde{\mathbf{K}}_{test} = (\mathbf{K}_{test} - \frac{1}{n}\mathbf{1}_{nt}\mathbf{1}_n^T\mathbf{K})(\mathbf{I} - \frac{1}{n}\mathbf{1}_n\mathbf{1}_n^T).
    \label{eq:centerNewKernel}
\end{equation}

The last step is making predictions for the new data

\begin{equation}
    \hat{\mathbf{Y}}_{test} = \tilde{\mathbf{K}}_{test} \mathbf{B}.
    \label{eq:predicitionMaking}
\end{equation}

\subsection{Kernel Partial Least-Squares with Kernel Flows}\label{ssec:KPLS-KF}

Kernel Flows (Owhadi \& Yoo), 2019 \cite{owhadi2019kernel}  was proposed to learn the hyperparameters of Kernel Regression. The method, presented in Algorithm \ref{alg:KF}, randomly extracts a percentage $p_b\%$ of the samples at each iteration. This data is denoted minibatch, and the subscript \textit{b} applies for all matrices of the minibatch. A sub-batch is extracted from this minibatch, and the subscript \textit{s} is utilized to describe the matrices of the sub-batch. 
The goal of the   two-level subset sampling  is to ensure that the model uniformly fits all subsets of the data. Here K-PLS is applied to both the minibatch and the sub-batches. 

With the desired class of kernel functions, the minibatch is mapped into the reproducing Kernel Hilbert space, as described in Subsection \ref{ssec:functions}, utilizing the kernel parameters $ \theta_i$ at iteration $i$. K-PLS is then performed between the minibatch kernel matrix ($\mathbf{K}_b$) and the response variable or matrix ($\mathbf{Y}_b$), and the regression coefficients are used to calculate the minibatch norm ($norm_b$)

\begin{equation}
    norm_b = \mathbf{B}_b^T\mathbf{K}_b\mathbf{B}_b.
\end{equation}

The same calculation is done for the sub-batch, consisting of a pre-defined fraction $p_s\%$ of the minibatch data. A K-PLS model is fitted on the sub-bach kernel matrix, and the response variable, and the corresponding sub-batch norm ($norm_s$) is calculated 

\begin{equation}
    norm_s = \mathbf{B}_s^T\mathbf{K}_s\mathbf{B}_s.
\end{equation}

The norms are used to compute the loss function 

\begin{equation}
    \rho = 1 - \frac{norm_s}{norm_b}.
\end{equation}

This study achieved better results if the sub-sampling was repeated several times for an iteration. For a number of sub-samplings $j$  in an iteration $i$ , the average loss ($\bar{\rho}_i$) for the iteration is


\begin{equation}
    \bar{\rho}_i = \frac{\sum_{j = 1}^{n_s} \rho_j}{n_s}
\end{equation}

The average loss ($\bar{\rho}_i$) at iteration $i$ is then used to calculate the gradient of the loss with respect to the kernel parameters ($\theta_i$). A gradient is calculated and used in the updating function for the kernel parameters of the next iteration ($\theta_{i+1}$). The gradient is calculated using automatic differentiation. 

\begin{algorithm}[H]
    \caption{Kernel Flows}\label{alg:KF}
    \textbf{\textit{Data}}: input matrix ($\mathbf{X}$), response matrix ($\mathbf{Y}$), number of latent variables ($n_{LVs}$), number of iterations ($n_i$), number of sub-samplings ($n_s$), percentage of samples in a batch ($p_b$), percentage of samples in a sub-batch ($p_s$), momentum coefficient ($\alpha$), initial Kernel function parameters ($\theta_0$). \textbf{\textit{Output}}: updated Kernel function parameters ($\boldsymbol{\theta}$).
    \begin{algorithmic}[1]
    \For{$i = 1, ..., n_i$}
    \State $\mathbf{X}_b, \mathbf{Y}_b \gets sample(\mathbf{X}, \mathbf{Y}, p_b)$ \Comment{Sample $p_b$\% of the data, the minibatch.}
    \State $\mathbf{B}_b \gets  kpls(\mathbf{X}_b, \mathbf{Y}_b, n_{LVs}, \theta_i)$ \Comment{Compute batch regression coefficients[Alg.\ref{alg:KPLS}]}
    \State $norm_b \gets \mathbf{B}_b^T\mathbf{K}_b\mathbf{B}_b$ \Comment{Compute minibatch norms.}
        \For{$j = 1, ..., n_s$}
            \State $\mathbf{X}_s, \mathbf{Y}_s \gets subsample(\mathbf{X}_b, \mathbf{Y}_b, p_s)$ \Comment{Subsample $p_s$ of the data}
            \State $\mathbf{B}_s \gets kpls(\mathbf{X}_s, \mathbf{Y}_s, n_{LVs}, \theta_i)$ \Comment{Compute sub-batch regression coefficients[Alg.\ref{alg:KPLS}]}
            \State $norm_s \gets \mathbf{B}_s^T\mathbf{K}_s\mathbf{B}_s$ \Comment{Compute sub-batch norms.}
            \State $\rho_j \gets 1 - \frac{norm_s}{norm_b}$ \Comment{Compute loss.}
        \EndFor
    \State $\bar{\rho}_i \gets \frac{\sum_{j = 1}^{n_s} \rho_j}{n_s}$ \Comment{Average loss.}
    \State $\nabla_{\theta_i} \gets diff(\bar{\rho}, \theta_i) $ \Comment{Compute gradients. }
    \State $\theta_{i+1} = update(\nabla_{\theta_i}, \theta_i, \alpha)$ \Comment{Update parameters for next iterations.}
    \EndFor     
    \end{algorithmic}
\end{algorithm}

In the original update function \cite{Owhadi2019}, $\theta_t$ is the parameter to be updated, $\alpha$ represents the learning rate and $\nabla_{\pmb \theta} f(\theta_t)$ is the gradient of the previous parameters' evaluation of the average loss function

\begin{equation}
    \theta_{t+1} = \theta_{t} - \alpha \nabla_{\pmb \theta} f(\theta_t).
    \label{eq:originalUpdate}
\end{equation}

Another version of parameter update is the Polyak momentum update \cite{sutskever2013importance}, where $\mu (\theta_t - \theta_{t-1}$ is a physics-inspired momentum in which $\mu$ is a hyperparameter usually in the range [0, 1]

\begin{equation}
    \theta_{t+1} = \theta_{t} - \alpha \nabla_{\pmb \theta} f(\theta_t) + \mu (\theta_t - \theta_{t-1}).
    \label{eq:polyak}
\end{equation}

The parameter update using Nesterov's \cite{su2014differential} momentum utilizes a momentum hyperparameter $\gamma$ that quantifies how much of the previous changes are to be added 

\begin{equation}
    \theta_{t+1} = \theta_{t} + \mu (\theta_t - \theta_{t-1}) - \gamma \nabla_{\pmb \theta} f(\theta_t + \mu (\theta_t - \theta_{t-1})).
    \label{eq:nesterov}
\end{equation}

The kernel parameters ($\boldsymbol{\theta}$) corresponding to the minimum loss value are considered as the optimized parameters. In some cases, the parameter value of the last iteration is saved. In other cases, a moving average of the loss vector is utilized. 

The optimal latent K-PLS variable dimension can be optimized in an outer loop, either by line-search or cross-validation. 

\subsection{Data division, pretreatment and metrics}\label{ssec:pretreatment}

For all the case studies, data has been split into 80 \% calibration and 20 \% testing partitions. The data has been standardized to zero-mean centres and unit-length variation. The calibration partition is denoted with the subscript \textit{cal}, whereas the test partition is denoted with the subscript \textit{test}.

To measure the success of the classification or regression tasks, the  following metrics were employed: 
\begin{itemize}
    \item \textit{Classification Accuracy} (CA):
        \begin{equation}
            CA = \frac{True Predictions}{Total Predictions}
        \end{equation}
    \item \textit{Root mean square error}, (RMSE):
        \begin{equation}
            RMSE = \sqrt{\frac{\sum_{i = 1}^n |\mathbf{Y}_{i, test} - \hat{\mathbf{Y}}_{i, test}|^2}{n}}
        \end{equation}
    \item \textit{Normalized Root Mean Square error} (NRMSE):
        \begin{equation}
            NRMSE = \frac{RMSE}{\mathbf{Y}_{max, cal} - \mathbf{Y}_{min, cal}} \cdot 100 
        \end{equation} 
    \item \textit{Goodness of prediction} ($Q^2$)
           \begin{equation}
               Q^2 = \frac{\sum_{i = 1}^n |\mathbf{Y}_{i, test} - \hat{\mathbf{Y}}_{i, test}|^2}{\sum_{i = 1}^n |\mathbf{Y}_{i, cal} - \bar{\mathbf{Y}}_{i, cal}|^2}
           \end{equation}
\end{itemize}

NRMSE is included for direct comparison with other K-PLS approaches in the literature.

\section{Case Studies}

This section presents the different case study datasets and their modelling goals. The modelling goals are either the (i) regression of a response variable for cases \ref{ssec:case1}, \ref{ssec:case3}, and \ref{ssec:case4}, or (ii) classification of samples. For the case studies, KF-PLS performance is compared to the original PLS, to K-PLS un-optimized and with K-PLS optimized with other optimization algorithms. 

\subsection{Case 1: Numerical example}
\label{ssec:case1}

The dataset has two synthetic variables having 200 observations, randomly generated between [-2, 2]. The mapping function is calculated as:

\begin{equation} \label{dataCase}
\begin{split}
f(\mathbf{x}_1, \mathbf{x}_2) = 3 (1-\mathbf{x}_1)^2 e^{\left(-\mathbf{x}_1^2 - (\mathbf{x}_2+1)^2 \right)} -  \\ 
10 \left( \frac{\mathbf{x}_1}{5} - \mathbf{x}_1^3 - \mathbf{x}_2^5 \right) e^{\left(-\mathbf{x}_1^2 - \mathbf{x}_2^2\right)}  -  \frac{1}{3} e^{\left( - (\mathbf{x}_1 + 1)^2 - \mathbf{x}_2^2\right)}.
\end{split}
\end{equation}

To the response variable, random noise is added of a chosen intensity \( \delta \):

\begin{equation} \label{dataNoisy}
\mathbf{y}^\delta = f(\mathbf{x}_1, \mathbf{x}_2) + \delta.
\end{equation}

\begin{figure}[H]
    \centering
    \includegraphics[width=0.8\linewidth]{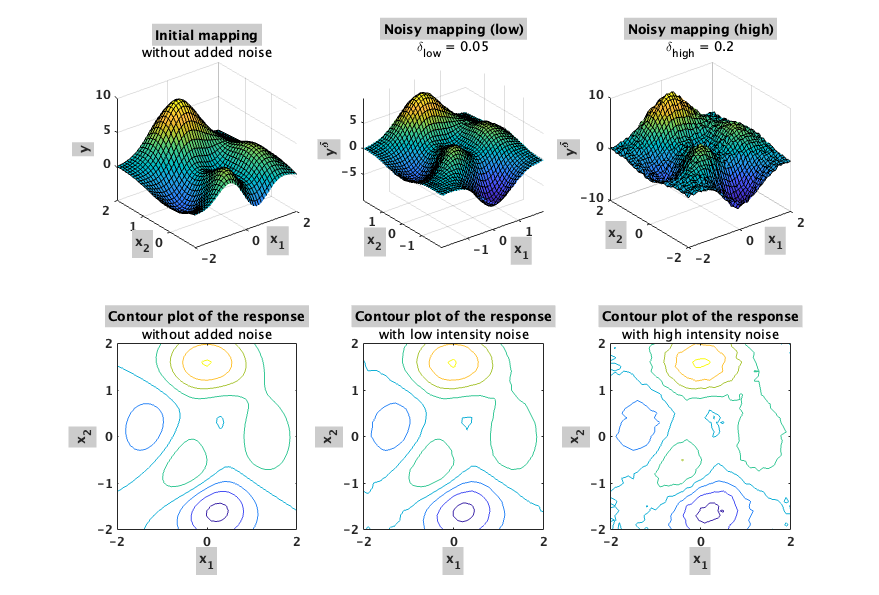}
    \caption{Surface and contour plots of the original mapping function, low-noise mappings \( \delta_{low} \) = 0.05 and high-noise mappings \( \delta_{high} \) = 0.2} 
    \label{fig:initialData}
\end{figure}

For illustration purposes, in this report, the noise level 0.05 will represent the low noise level, and results associated with it will be denoted with \( \delta_{low}\), whereas the noise level 0.2 will represent a high noise level and will be denoted with \( \delta_{high} \). Fig. \ref{fig:initialData} shows the original function with no noise and its low and high \( \delta \) noisy versions. 

The purpose of this example is to estimate $\mathbf{y}$ from $\mathbf{x}_1$ and $\mathbf{x}_2$ without fitting noise into the K-PLS model. The evaluation of the predicted $\hat{\mathbf{y}}$ is done with the mapped $\mathbf{y}$ as reference. 

\subsection{Case 2: Classification of concentric circles}\label{ssec:case2}

\begin{figure}[H]
    \centering
    \includegraphics[width=0.5\textwidth]{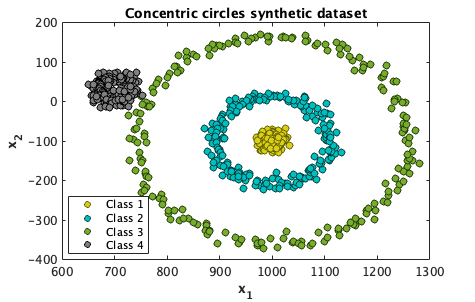}
    \caption{Concentric circles dataset used for non-linear classification task.}
    \label{fig:concCenters}
\end{figure}

The following case study captures a non-linear classification scenario where concentric circles are represented in two dimensions. The dataset is showcased in Figure \ref{fig:concCenters}. The case study is a non-linear classification task, as no line can be fitted in the original two dimensions to separate the classes. 

K-PLS with Discriminant Analysis (K-PLS-DA) aims to classify the data into four classes. To perform the classification, a dummy variable is created for each class. The value for the sample that takes membership in a class is 1 for the given class and 0 for the others. The resulting response matrix has four variables, one for membership in each class.

\subsection{Case 3: Predicting the strength of concrete}\label{ssec:case3}

\begin{figure}[H]
    \centering
    \includegraphics[width=\linewidth]{ 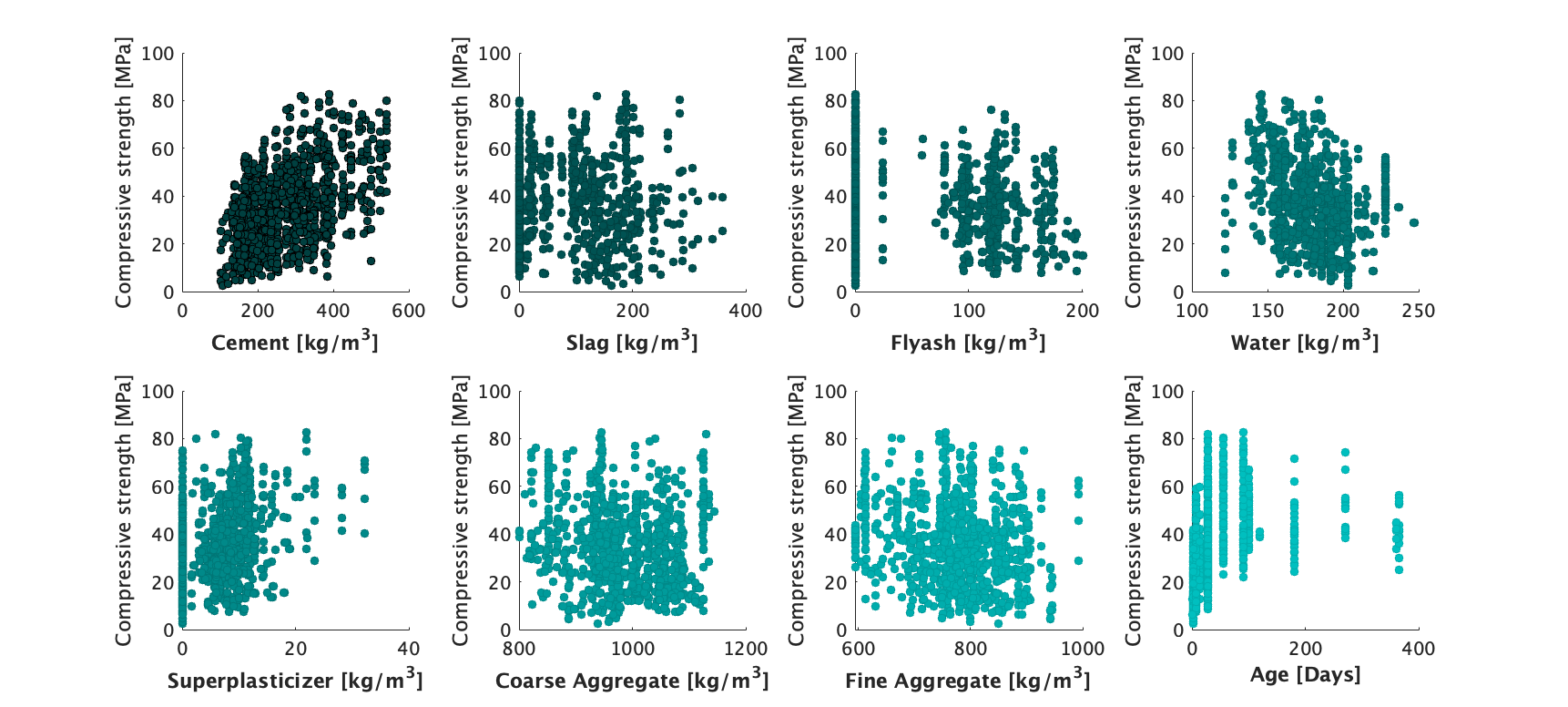}
    \caption{Compressive strength dataset: individual variables against the response variable.}
\label{fig:cement}
\end{figure}

A commonly met scenario of non-linear dependencies between the available data and the response variable is found in the concrete strength estimation from compositional and age information. Figure \ref{fig:cement} presents the dataset \cite{yeh1998modeling}. The prediction of concrete strength - the response variable - relies on the underlying non-linear dependency between cement, slag, fly ash, water, superplasticiser and coarse aggregate contents $[kg/m^3]$, along with the age of the concrete measured in days. 

\subsection{Case 4: Soft sensor model for soil moisture prediction from hyperspectral data}\label{ssec:case4}

In the previous scenarios, the original data variable space is likely to be lower than the dimensions in which the dependencies are linear. A hyperspectral dataset has been used to test the efficiency of KF-PLS in the opposite scenario, where the original space is already high-dimensional. This dataset contains soil spectra and soil moisture content for each spectrum. The dataset is a benchmark \cite{riese2018hyperspectral}, and the modelling aim is to create a soft sensor model to estimate the soil moisture from hyperspectra.

\begin{figure}[H]
    \centering
    \begin{subfigure}[b]{0.23\linewidth}
        \centering
        \includegraphics[width=0.8\linewidth]{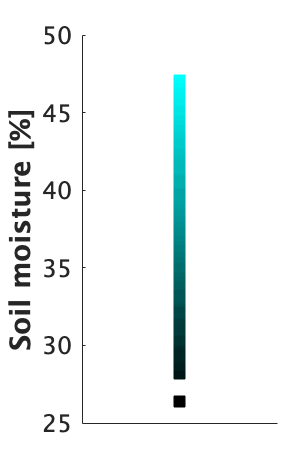}
    \end{subfigure}
    \begin{subfigure}[b]{0.65\linewidth}
        \centering
        \includegraphics[width=0.85\linewidth]{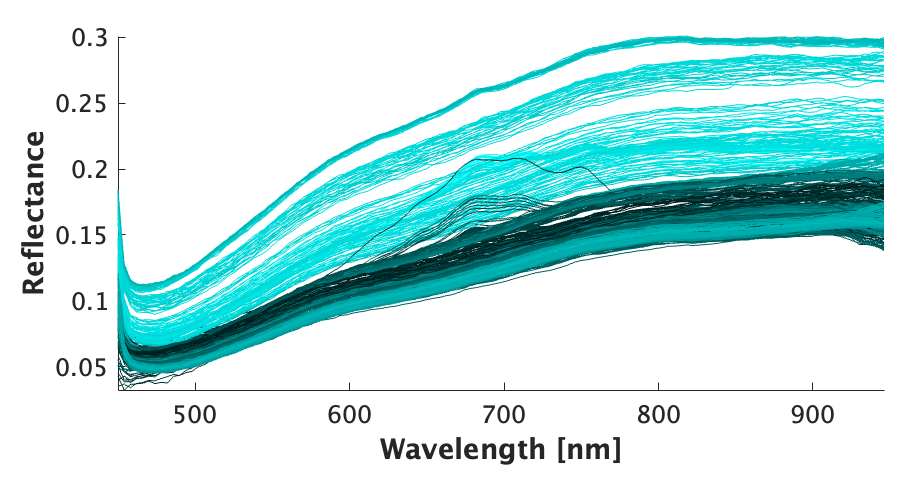}
    \end{subfigure}
    \caption{Hyperspectral benchmark dataset for soil moisture prediction.}
    \label{fig:hyperspectralData}
\end{figure}

\section{Results}

Table \ref{tab:baseResults} presents the main results for the four case studies. The results show that KF-PLS is able to enhance regression and classification results in all case studies. More detailed analysis is provided in the sub-sections dedicated to the sensitivity studies for each case.

\begin{table}[H]
    \centering
    \caption{General case studies results in optimized KF-PLS, compared to the regular PLS. The performance was improved in all cases. *The number of iterations until convergence. $\uparrow$ indicates if the metric is optimal if maximized, whereas $\downarrow$ indicates that indicator is to be minimized. }
    \begin{tabular}{c|c|c|c|c|c|c|c}
        \textbf{Case} & \multicolumn{3}{c|}{\textbf{Original PLS}} & \multicolumn{4}{c}{\textbf{Kernel-Flows Optimized (KF-PLS)}} \\
         & $Q^2_{test} \uparrow $ & NRMSE $\downarrow$ & Accuracy $\uparrow$ & $Q^2_{test} \uparrow $ & NRMSE $\downarrow$  & Accuracy $\uparrow$  & Iterations*\\
         \hline
         1 & 0.85 & 5.2 & - & 0.99 & 1.21 & - & 213 \\
         2 & -  & - & 0.7 & - & - & 1 & 70 \\
         3 & 0.9 & 0.82 & - & 0.96 & 0.53 & - & 30 \\
         4 & 0.94 & 2.36 & - & 0.97 & 1.68 & - & 65\\
    \end{tabular}
    \label{tab:baseResults}
\end{table}

Apart from their case-specific modelling goal, each case study also serves as a sensitivity study for KF-PLS. The particular goals and kernel functions utilized in this case study are presented in Table~\ref{tab:caseStudies}.


\begin{table}[H]
    \centering
    \caption{Characteristics and sensitivity studies conducted in each of the test cases.}
    \begin{tabular}{c|c|c|c}
        \textbf{Case} & \textbf{Data type} & \textbf{Kernel function}  & \textbf{Sensitivity studies}\\
        \hline
        1 & Synthetic & Gaussian & Noise effects \\
          & & Gaussian & Latent variables study \\
          & & All & Influence of different kernel functions \\
          & & All & Influence of the number of sub-samplings \\
        \hline
        2 & Synthetic & Gaussian  & Influence of initial kernel parameters\\
          & & & Influence of initial learning rate \\
          & & & Comparison of different optimization methods \\
        \hline
        3 & Real & Cauchy & Optimizing the latent variables \\
        \hline
        4 & Real & Gaussian & Effects of data scaling \\
    \end{tabular}
    \label{tab:caseStudies}
\end{table}

\subsection{Case 1: Noise effects and kernel function studies}

The answer to the research question "\textit{Is fine-tuning the K-PLS parameters with KF overfitting the training data}?" is explored within Case Study 1. 

The models have been calibrated with noisy data (Eq. \ref{dataNoisy}), and the test partition results are compared to the (a) true noiseless underlying mapping (Eq. \ref{dataCase}) and the (b) noisy mapping to demonstrate that KF-PLS models capture the underlying structure of the data and not the random noise added. The KF-PLS predictions ($\hat{\textbf{y}}$) are compared in the RMSE indicator to the true mapping, $RMSE(\hat{\textbf{y}}, f(\textbf{X}))$ and to the noisy mapping $RMSE(\hat{\textbf{y}}, f(\textbf{X})+\delta)$.

\begin{figure}[H]
    \centering
    \begin{subfigure}[b]{0.55\linewidth}
        \centering
        \includegraphics[width=0.7\linewidth]{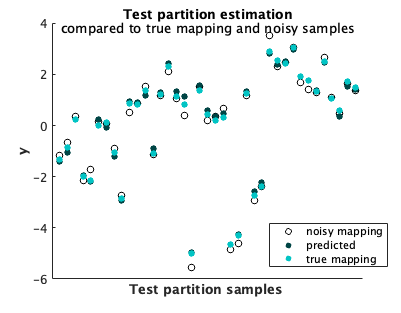}
        \caption{The de-noising effect of KF-PLS on the test partition prediction for Case Study 1 with a high level of noise $\delta_{high} = 0.2$. The figure shows that the residuals are closer to the true mapping of the function, than the noisy mapping of the function, and that the residuals are not varying with the y-value increase.}
        \label{fig:denoising}
    \end{subfigure}
    \hfill
    \begin{subfigure}[b]{0.42\linewidth}
        \centering
        \includegraphics[width=\linewidth]{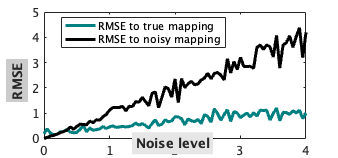}
        \caption{RMSE to the original mapping and noisy data, for the test partition, in different noise intensities, where the noise level is the $\delta$ parameter in. Eq. \ref{dataNoisy}. The figure shows that, as the noise level in the calibration partition increases, the model predictions for the test data stay true to the original mapping, and the distance to the noisy mapping increases proportionally to the increase in noise level.}
        \label{fig:noiseEffects}
    \end{subfigure}
   \caption{Case 1: Noise effect studies on non-linear regression.}
   \label{fig:noise} 
\end{figure}

As seen in Figure \ref{fig:denoising}, the predicted samples from the test partition are closer to the true function mapping, rather than the noisy data used in calibration. The results are also confirmed in Figure \ref{fig:noise}, where increasing the noise level $ \delta $ in Eq. \ref{dataNoisy}, the RMSE between the prediction and the noisy data increases almost linearly, whereas the RMSE between the K-PLS estimation and the true function mapping remains almost at the same level. This result is achieved due to the power of PLS-based methods to filter out non-representative variance, such as noise from the model, by selecting a proper number of latent variables; thus, the included variance. 

\begin{figure}[H]
    \centering
    \includegraphics[width=0.6\linewidth]{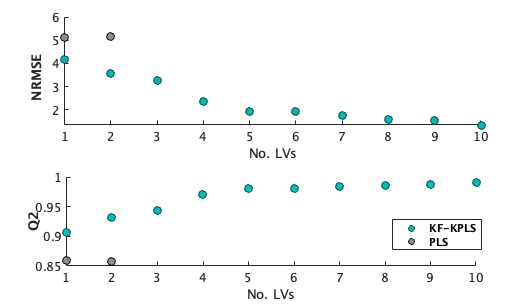}
    \caption{Outer loop evaluation of the K-PLS number of dimensions, compared to PLS. K-PLS can be represented in higher dimensions than the original two-dimensional space, and in this case, it is an advantage, as the optimum results occur around 4 LVs. After 4 LVs, the improvement is shallow, and the complexity addition to the model is not sustained.}
    \label{fig:LVscase1}
\end{figure}

The effect of the number of latent variables on regression performance is shown in Figure. \ref{fig:LVscase1}.  The evaluated results are calculated for the test partition, which has not been included in the KF optimization or the PLS model calibration. The maximum amount of latent variables available for the original PLS is two, as the dataset only has two predictor variables. By extending the variable space by projecting the data using the kernel functions, the increased number of latent variables seems to benefit the prediction of the independent samples, with the performance increasing significantly: more than  14 \% of the $Q^2$  value achieved with the regular PLS.

\begin{figure}[H]
    \centering
    \includegraphics[width=0.4\linewidth]{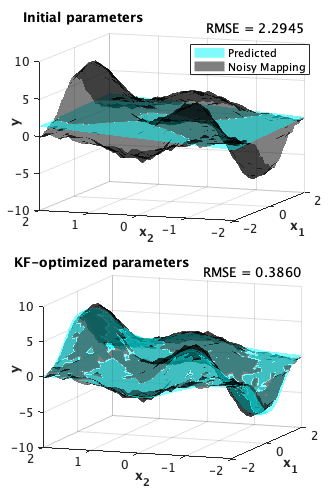}
    \caption{Overlay of response surfaces in case of the noisy data and KF-PLS estimation. The predicted feature is not well represented in the initial case, where the hyperparameters of the Gaussian kernel are '1': the $\mathbf{y}$-prediction is a plane located at the average value of $\mathbf{y}$. The model with KF-optimized parameters is able to represent correctly the $\mathbf{y}$-shape, without overfitting on the noise and with correct estimation of the extreme values.}
    \label{fig:surfaceplots}
\end{figure}

KF-PLS can find the optimal Kernel parameters, regardless of the initializing values for the Kernel Function parameter. For example, in Figure \ref{fig:surfaceplots}, the convergence of the Gaussian Kernel parameters can be observed from very far-off initial parameters to parameters that allow for an accurate replica of the mapping function without over-fitting the noise. 

\begin{figure}[H]
    \centering
    \begin{subfigure}[b]{0.49\linewidth}
        \centering
        \includegraphics[width=\linewidth]{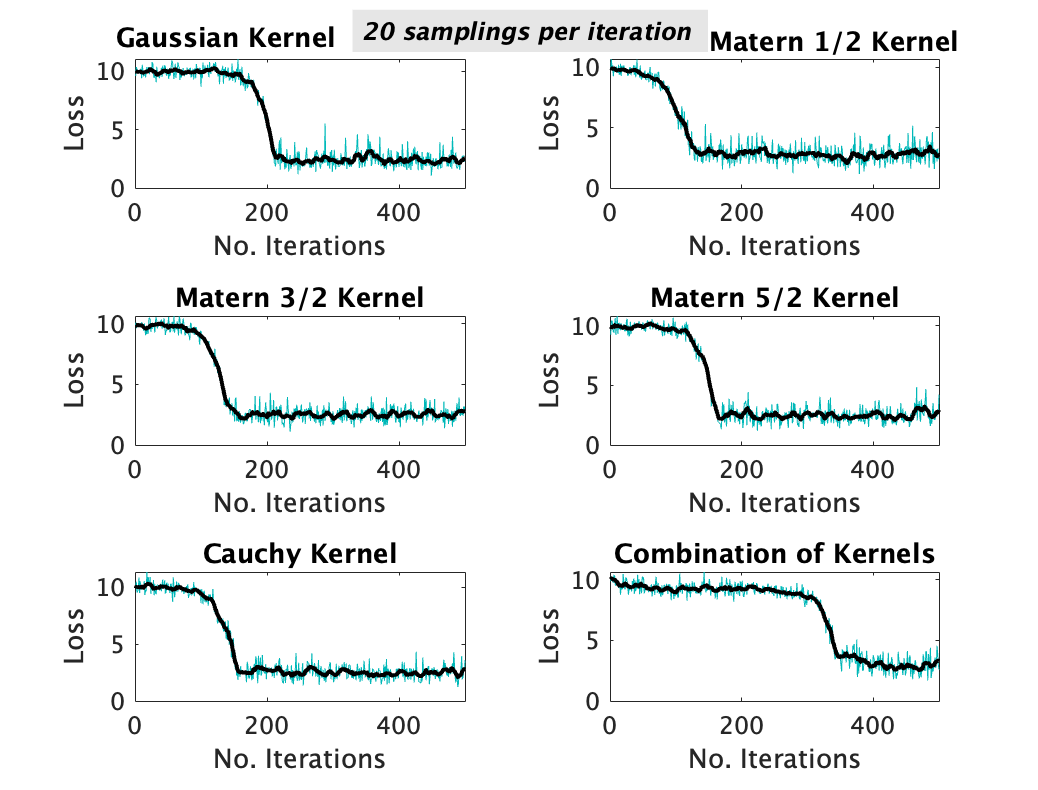}
        \caption{}
        \label{fig:lossAverage}
    \end{subfigure}
    \hfill
    \begin{subfigure}[b]{0.49\linewidth}
        \centering
        \includegraphics[width=\linewidth]{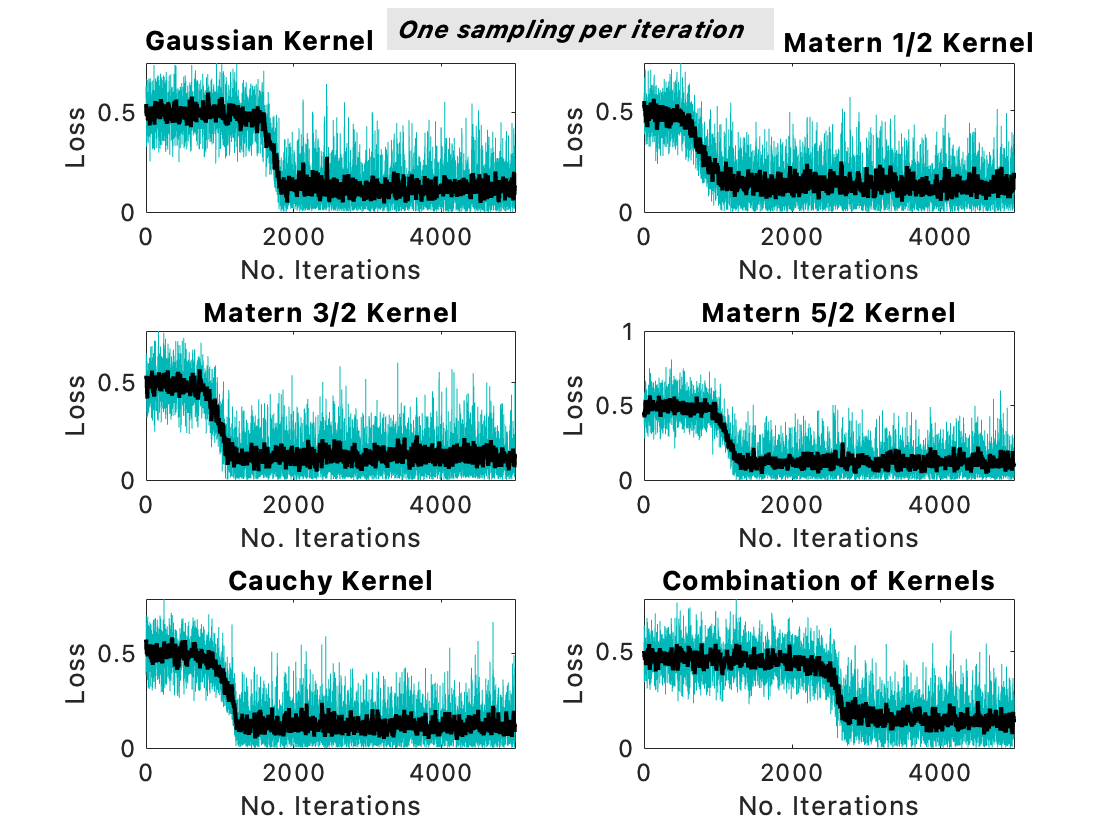}
        \caption{}
        \label{fig:lossRaw}
    \end{subfigure}
    \caption{Loss function values for 20 sub-samplings per iteration (a) compared to one sub-sampling per iteration (b). Note that the scale in (b) is 10 times larger. When there is only one sub-sample per iteration, the optimisation takes more time to converge and the loss is less stable. When there are 20 sub-samplings per iteration, the loss function is more stable. When a combination of kernels is utilized, both the single sub-sample and multiple sub-sample cases converge slower.}
    \label{fig:losses}
\end{figure}

Convergence of Kernel Flows depends on the properties of each data set and the number of sub-samplings for each iteration. In Figure \ref{fig:losses}, two scenarios are presented: the scenario in which multiple sub-samplings have been employed to calculate the loss function and the scenario in which the loss function represents a single sub-sampling per iteration. Including more sub-samplings gives more stability, reflected in the iteration-to-iteration variation, and a shorter number of iterations until convergence. 

Datasets with local non-linearities and an uneven sample distribution across the variation areas benefit from the sub-sampling. For example, the peaks in the function or extreme values can have a poorer representation in the dataset. Thus, the chances of these points being sub-sampled in an iteration are lower. Also, K-PLS is truncating the data variation to a fixed number of latent variables. If a subsample does not represent the whole dataset, the same number of latent variables can include a different amount of variation, possibly fitting noise. Increasing the number of subsamples per iteration ensures a more accurate error metric for the iteration.  

\subsection{Case 2: Effects of learning rate and initial parameter values on  converge of Kernel Flows}

\begin{figure}[H]
    \centering
    \begin{subfigure}[b]{0.49\linewidth}
        \centering
        \includegraphics[width=1\linewidth]{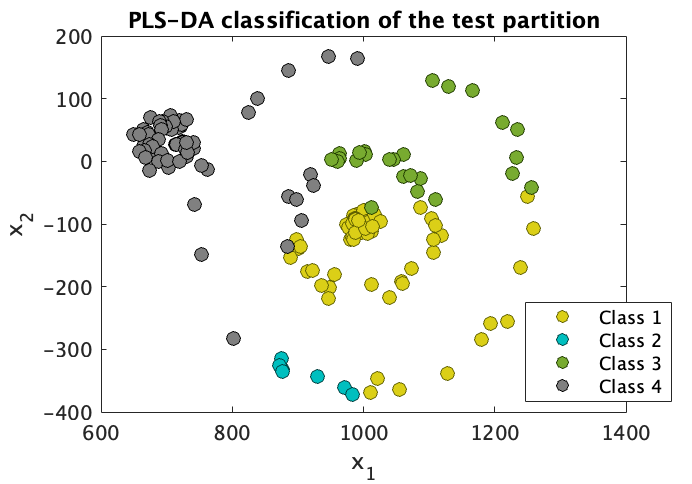}
        \caption{}
        \label{fig:resultsCase2PLSDA}
    \end{subfigure}
    \hfill
    \begin{subfigure}[b]{0.49\linewidth}
        \centering
        \includegraphics[width=1\linewidth]{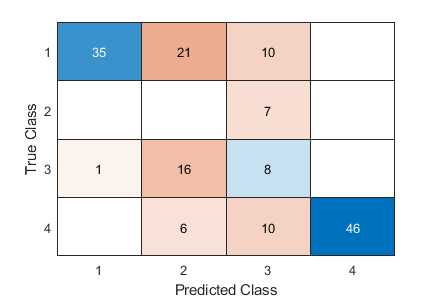}
        \caption{}
        \label{fig:confusionCase2PLSDA}
    \end{subfigure}
    \hfill
    \begin{subfigure}[b]{0.49\linewidth}
        \centering
        \includegraphics[width=1\linewidth]{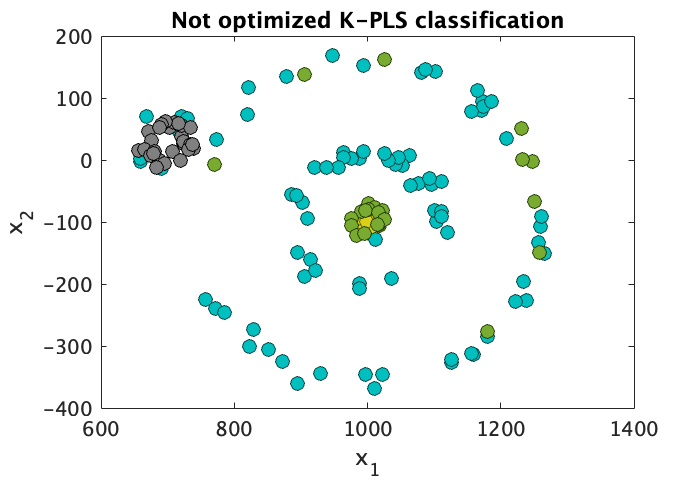}
        \caption{}
        \label{fig:resultsCase2Unoptimized}
    \end{subfigure}
    \hfill
    \begin{subfigure}[b]{0.49\linewidth}
        \centering
        \includegraphics[width=1\linewidth]{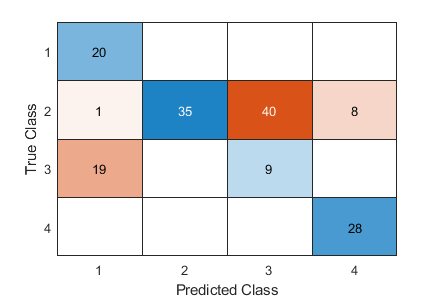}
        \caption{}
        \label{fig:confusionCase2Unoptimized}
    \end{subfigure}
    \hfill
    \begin{subfigure}[b]{0.49\linewidth}
        \centering
        \includegraphics[width=1\linewidth]{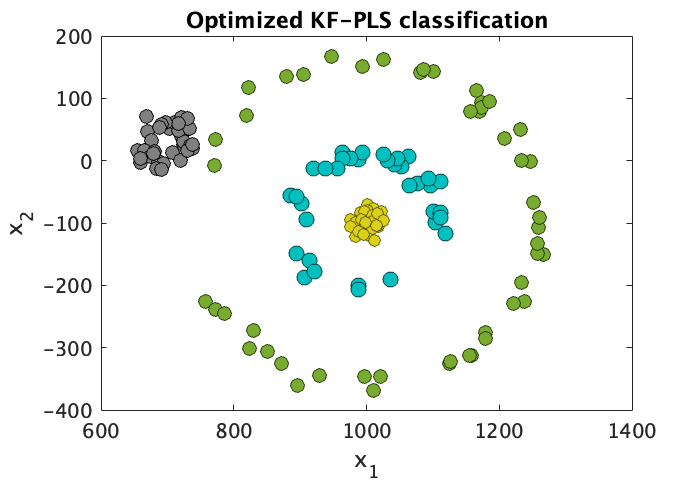}
        \caption{}
        \label{fig:resultsCase2Optimized}
    \end{subfigure}
    \hfill
    \begin{subfigure}[b]{0.49\linewidth}
        \centering
        \includegraphics[width=1\linewidth]{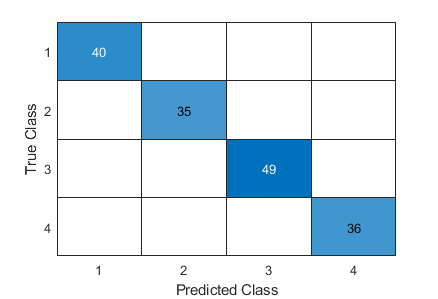}
        \caption{}
        \label{fig:confusionCase2Optimized}
    \end{subfigure}
    \caption{Classification results of the test partition with PLS-DA (a-b), un-optimized K-PLS (c-d) and optimized KF-PLS (e-f). The prediction accuracy for PLS-DA is 55\%, for un-optimized K-PLS 57\% and for KF-PLS 100\%.}
    \label{fig:resultsClassification}
\end{figure}

As a bi-linear method, PLS-DA cannot separate between the concentric circles and correctly classify. As seen in Figures \ref{fig:resultsCase2PLSDA},  three out of the four classes are incorrectly classified. When using K-PLS, different concentric circles can be allocated to separate classes. The non-optimized K-PLS gives slightly better results than the regular PLS-DA, but according to Figure \ref{fig:resultsCase2Unoptimized} the accuracy is still low. The KF-optimized classification, with three latent variables in K-PLS, leads to a perfect classification of the test partition as seen in Fig. \ref{fig:resultsCase2Optimized} and \ref{fig:confusionCase2Optimized}. 

\begin{figure}[H]
    \centering
    \includegraphics[width=1\linewidth]{ 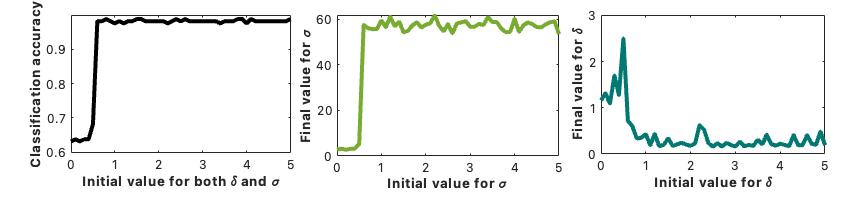}
    \caption{Classification accuracy using the convergence parameters for 500 iterations and a Gaussian kernel. $\sigma$ and $\delta$ are kernel parameters from Eq. \ref{eq:gaussianKernel} and Eq. \ref{eq:regularizaion}, and their initial value is equal ($\sigma_{initial}$ = $\delta_{initial}$). When 500 iterations are utilized, models with initial values > 0.6 converge. For small initial parameter values, more iterations are needed. If the algorithm converges, the final value for the parameters at the convergence time is similar, regardless of the starting value of the parameters.}
    \label{fig:convergenceC2}
\end{figure}

The initialization value for the Kernel parameters has proven not to be important for the final result. Every time Kernel Flows reaches convergence, the parameters settle around the same value. Figure \ref{fig:convergenceC2} shows an example where a Gaussian kernel has been used, and KF was run with 500 iterations. The algorithm has not yet converged in 500 iterations for the low initial values, so the final parameters are not settled around the same numerical value. The final convergence value is similar when the algorithm converges, regardless of the initial value for the parameters.  

\begin{figure}[H]
    \centering
    \includegraphics[width=0.5\linewidth]{ 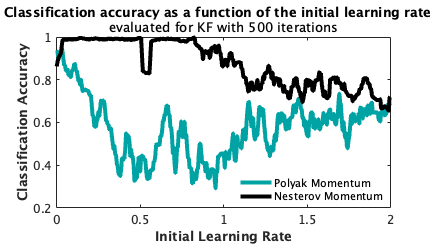}
    \caption{Effect of the initial learning rate on classification accuracy for a Gaussian kernel and 500 iterations. For small learning rates, both momentums have similar performance. The Nesterov momentum performs best when optimizing the parameters of a single Kernel. Polyak's momentum has the best performance when there is a combination of kernels to be optimized. The results represent an average of 15 separate experiments.}
    \label{fig:momentumAndRateC2}
\end{figure}

The updating momentum's effect on the convergence time is also investigated. For 500 iterations, different initial learning rates and the classification accuracy of the test partition are presented in Figure \ref{fig:momentumAndRateC2}. Experimental observations show that Polyak's momentum needs longer optimization runs until convergence for lower initial learning rates, whereas Nesterov's momentum works best with low initial learning rates. Nesterov's momentum has been proven more efficient in cases where there are many kernel parameters to be optimized, for example, in the combination of kernels.

\begin{figure}[H]
    \centering
    \includegraphics[width=0.7\linewidth]{ 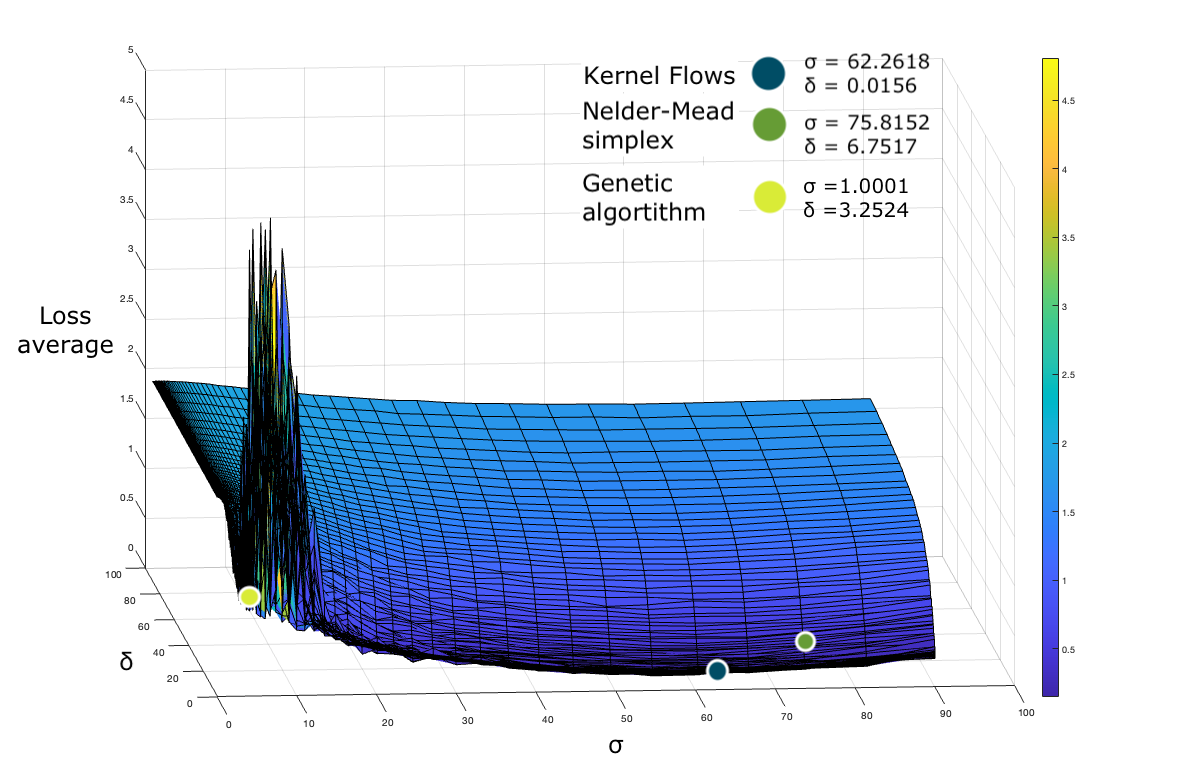}
    \caption{Loss map using different parameters for regularized Gaussian kernel. The genetic algorithm is an Augmented Lagrangian Genetic Algorithm. While the genetic algorithm and Nedler-Mead simplex land in 
    local minima, KF-PLS correctly converges in the global minimum of the loss function.}
    \label{fig:lossMaps}
\end{figure}

The Kernel Flows algorithm is able to converge to the global loss minimum. For comparison, the same loss function was minimized using different optimization methods: the Augmented Lagragian Genetic Algorithm, and Nelder-Mead simplex. The loss surface rendered in Figure \ref{fig:lossMaps} includes the average loss function values, where for each combination of the Gaussian Kernel parameters $\sigma$ and $\delta$, five repeating iterations with twenty averaged subsamplings for each iteration were considered. The analysis revealed that Kernel Flows using Stochastic Gradient Descent was the fastest algorithm to converge to the minimum and had the lowest loss function for the convergence point, compared to the Genetic Algorithm or the Nedler-Mead simplex. 

\subsection{Case 3: The effect of initial number of latent variables}

\begin{figure}[H]
    \centering
    \begin{subfigure}[b]{0.65\linewidth}
        \centering
        \includegraphics[width=0.8\textwidth]{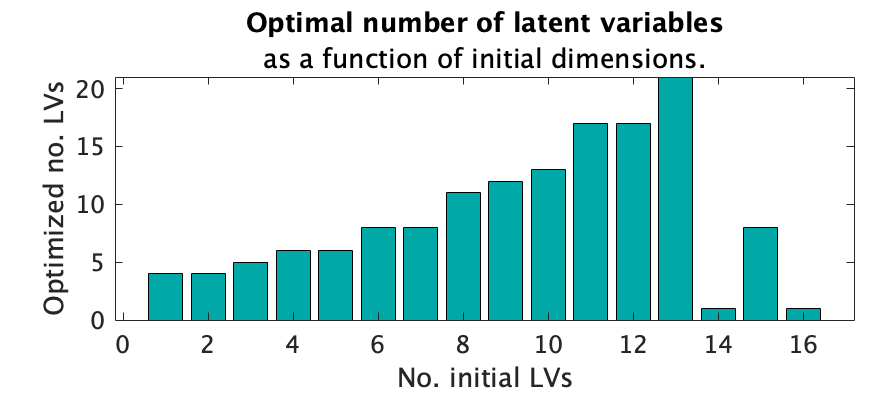}
        \caption{}
        \label{fig:C4finalLVs}
    \end{subfigure}
    \hfill
    \begin{subfigure}[b]{0.65\linewidth}
        \centering
        \includegraphics[width=0.8\textwidth]{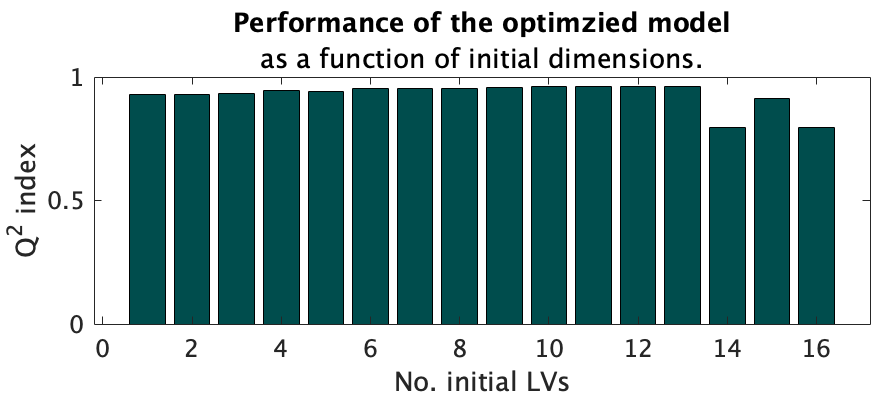}
        \caption{}
        \label{fig:C4q2}
    \end{subfigure}
    \hfill
    \begin{subfigure}[b]{0.65\linewidth}
        \centering
        \includegraphics[width=0.8\textwidth]{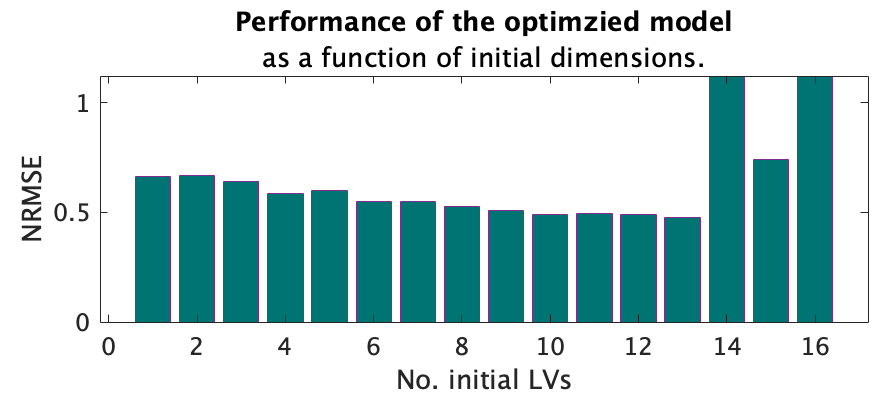}
        \caption{}
        \label{fig:C4nrmse}
    \end{subfigure}
    \caption{Studies of initial latent variables effect on prediction results of the test partition. The initial decision on the number of latent variables does not significantly affect the quality of results (similar $Q^2$ and NRSME). When the number of LVs is too big, the KF no longer converges. If too many latent variables are taken from a small-size sub-batch, data is over-fitted, and the loss function is unstable.}
    \label{fig:case4figureLVs}
\end{figure}

The number of optimal latent variables can be set or checked in an outer optimization loop. As shown in Figure \ref{fig:C4finalLVs}, the optimal number of latent variables is often slightly higher with one to two LVs than the starting point. Still, there is no significant difference in the evaluation of the test partition metrics when looking at the $Q^2$ scores in Figure \ref{fig:C4q2}. 

When including too many latent variables, noise or variance not related to the modelling task is fitted, thus decreasing the performance of the model. One notable observation has to do with the size of the sub-sampling batch. If the subsampling batch has a number of \textit{i} samples, the maximum dimension possibly achieved is \textit{i} latent variables. This number represents 100 \% of the co-variation in the subsampling batch, including the possible noise variation desired to be filtered out and equivalent to multiple linear regression. As a rule of thumb, doing the Kernel Flows optimization with fewer latent variables is recommended. This decreases overfitting while representing the variation in the data. After setting the kernel parameters, a line search can optimise the number of latent variables.

\subsection{Case 4: Effect of data scaling}

The case study illustrates the efficiency of KF-PLS, in the case where the underlying structure is represented in fewer dimensions than the original data. In Case 2, the dimensions have increased past the original dimensions, and the dependency is linear in a higher dimension. In Case 4, the opposite happens: hyperspectral data has plenty of dimensions to begin with.  

\begin{figure}[H]
    \centering
    \begin{subfigure}[b]{0.45\linewidth}
        \centering
        \includegraphics[width=\linewidth]{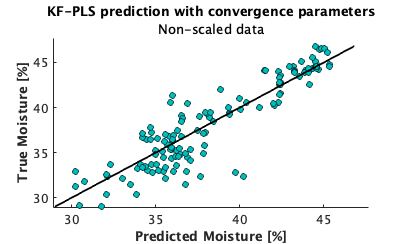}
        \caption{}
        \label{fig:unscaledResults}
    \end{subfigure}
    \hfill
    \begin{subfigure}[b]{0.45\linewidth}
        \centering
        \includegraphics[width=\linewidth]{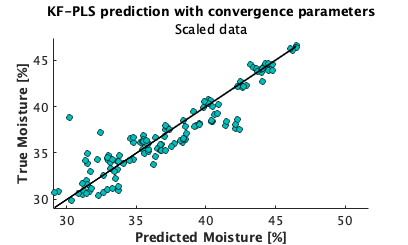}
        \caption{}
        \label{fig:scaledResults}
    \end{subfigure}
    \hfill
    \begin{subfigure}[b]{0.45\linewidth}
        \centering
        \includegraphics[width=\linewidth]{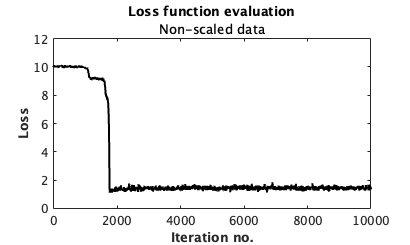}
        \caption{}
        \label{fig:lossUnscaled}
    \end{subfigure}
    \hfill
    \begin{subfigure}[b]{0.45\linewidth}
        \centering
        \includegraphics[width=\linewidth]{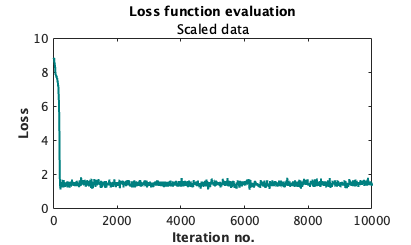}
        \caption{}
        \label{fig:lossScaled}
    \end{subfigure}
    \hfill
    \begin{subfigure}[b]{0.45\linewidth}
        \centering
        \includegraphics[width=\linewidth]{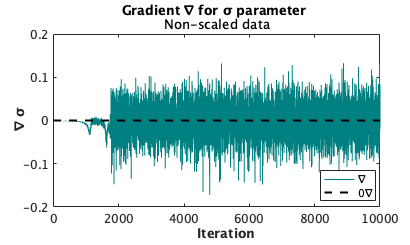}
        \caption{}
        \label{fig:gradSigmaUncaled}
    \end{subfigure}
    \hfill
    \begin{subfigure}[b]{0.45\linewidth}
        \centering
        \includegraphics[width=\linewidth]{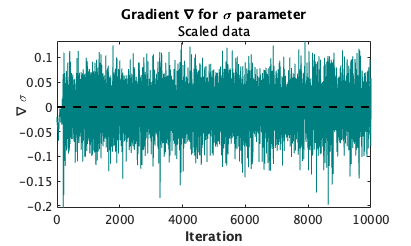}
        \caption{}
        \label{fig:gradSigmaScaled}
    \end{subfigure}
    \hfill
    \begin{subfigure}[b]{0.45\linewidth}
        \centering
        \includegraphics[width=\linewidth]{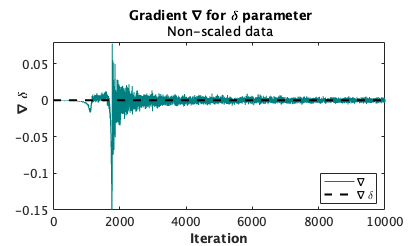}
        \caption{}
        \label{fig:gradDeltaUncaled}
    \end{subfigure}
    \hfill
    \begin{subfigure}[b]{0.45\linewidth}
        \centering
        \includegraphics[width=\linewidth]{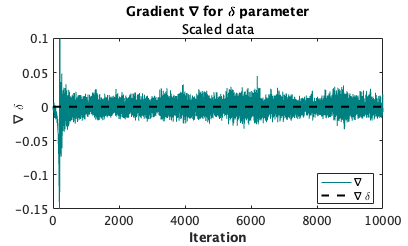}
        \caption{}
        \label{fig:gradDeltaScaled}
    \end{subfigure}
    \hfill
    \caption{Test partition results, differentiation gradients and loss function results for scaled and non-scaled data. In the image, "Gradient" denotes a gradient component.}
    \label{fig:case4}
\end{figure}

Figure \ref{fig:case4} represents a comparison between results in which data has not been centred and scaled and when it has been centred and scaled. The difference in results is not significant, but there is a significant difference in the convergence time. While the non-scaled version took close to 2000 iterations to converge (Figure \ref{fig:lossUnscaled}), the scaled version converged much faster when data was scaled and centred according to the methodology proposed in subsection \ref{ssec:pretreatment}. The gradient value for the kernel parameters are on the same scale, and the gradients are centred at 0 in the convergence time. 

\section{Conclusions}

In the present paper, we propose a method to learn the Kernel hyperparameters for Kernel-Partial Least-Squares (K-PLS) using Kernel Flows (KFs) methodology. The method has been proven to be efficient in learning the Kernel parameters, regardless of the Kernel mapping function, for both regression with K-PLSR or classification using K-PLS with Discriminant Analysis (K-PLS-DA).

The methodology has been tested with synthetic and real data to model non-linear regression or classification tasks in four case studies. The case studies showcased good performance for the resulting models that have been evaluated by applying accuracy, RMSE and $Q^2$ indicators on the testing partitions of the datasets. Both scenarios in which the number of initial dimensions of the data is increased or decreased are considered. The dimensionality in which linear dependencies exist between the predictor and response variables is found in both scenarios. 

The Kernel Flows algorithm requires the user to set the values of a number of hyperparameters. Apart from the PLS dimensions, which can be optimized in an outer evaluation loop, the case studies have proven that the algorithm converges to an optimal value regardless of the initial input of the parameters. This includes the initial values for the kernel parameters and the initial learning rate. Hyperparameters such as the number of sub-samplings are not critical for convergence, but they can influence the number of iterations until convergence and loss function stability. The usage of an updating momentum can also result in shorter convergence times. Thus, the KFs methodology is able to successfully learn the Kernel parameters for K-PLS, bringing optimal performance for non-linear regression and classification tasks while still maintaining the qualities of PLS: restricting the amount of relevant variation that the model considers and avoiding over-fitting.

\section{Acknowledgements}
Funding from Research Council of Finland for Centre of Excellence of Inverse Modelling and Imaging, project number 353095, is acknowledged.  The research effort by JS and OL was carried out at the Jet Propulsion Laboratory, California Institute of Technology, under a contract with the National Aeronautics and Space Administration (80NM0018D0004).

\end{document}